\documentclass[%
reprint,
superscriptaddress,
showpacs,preprintnumbers,
nofootinbib,
amsmath,amssymb,
aps,
prd,
floatfix,
]{revtex4-1}

\usepackage{graphicx}% Include figure files
\usepackage{dcolumn}% Align table columns on decimal point
\usepackage{bm}% bold math
\usepackage{hyperref}% add hypertext capabilities
\usepackage[mathlines]{lineno}% Enable numbering of text and display math
\usepackage{float}
\usepackage{tabulary}
\usepackage{color}

\begin{document}

%\preprint{}

\title{Astrophysical uncertainties on stellar microlensing constraints on multi-Solar mass primordial black hole dark matter}

\author{Anne M. Green}
\email{anne.green@nottingham.ac.uk}
\affiliation{School of Physics and Astronomy, University of Nottingham, University Park, Nottingham, NG7 2RD, United Kingdom}

\date{\today}

\begin{abstract}
There has recently been interest in multi-Solar mass Primordial Black Holes (PBHs) as a dark matter (DM) candidate. There are various  microlensing, dynamical and accretion constraints on the abundance of PBHs in this mass range. Taken at face value these constraints exclude multi-Solar mass PBHs making up all of the DM for both delta-function and extended mass functions. However the stellar microlensing event rate depends on the density and velocity distribution of the compact objects along the line of sight to the Magellanic Clouds. We study the dependence of the constraints on the local dark matter density and circular speed and also consider models where the velocity distribution varies with radius. We find that the largest mass constrained by stellar microlensing can vary by an order of magnitude. In particular the constraints are significantly weakened if the velocity dispersion of the compact objects is reduced. The change is not sufficiently large to remove the tension between the stellar microlensing and dynamical constraints.
However this demonstrates that it is crucial to take into account astrophysical uncertainties when calculating and comparing constraints. We also confirm the recent finding that the tension between the constraints is in fact increased for realistic, finite width mass functions.
\end{abstract}

\maketitle

\section{Introduction}

Primordial Black Holes (PBHs) can form in the early Universe from the collapse of large amplitude overdensities~\cite{Carr:1974nx,Carr:1975qj}. PBHs with mass $M \gtrsim 10^{15} \, {\rm g}$ are stable and since they form before nucleosynthesis are non-baryonic. PBHs are therefore a potential cold dark matter (DM) candidate. There are various constraints on their abundance, from gravitational lensing and their dynamical and other effects on various astrophysical objects and processes (see Ref.~\cite{Carr:2016drx} for a detailed compilation of the constraints as of mid-2016).

\begin{figure}[t]
%\begin{center}
\includegraphics[width=0.45\textwidth]{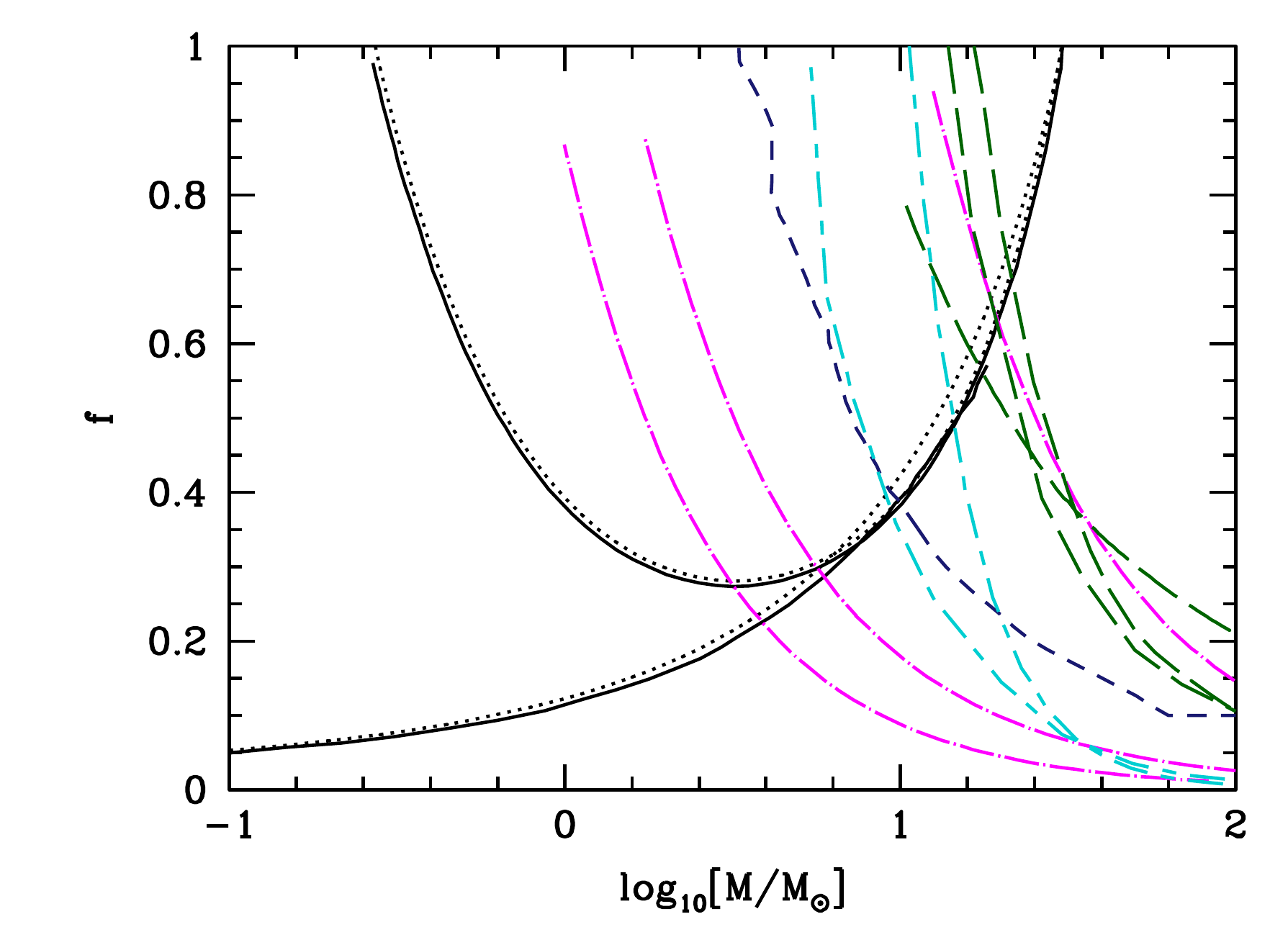}
%\end{center}
\caption{\label{fig:flimall} The constraints on the dark matter fraction, $f$, of PBHs with mass $M$  in the multi-Solar mass region, assuming a delta-function mass function. The constraints from LMC microlensing surveys are shown as solid black lines, with the corresponding dotted lines showing our reproduction of these limits, as described in Sec.~\ref{sec:micro}. 
The two sets of lines are, from top to bottom at small $M$, the MACHO collaboration $(1-30) M_{\odot}$ black hole search~\cite{Allsman:2000kg} and the EROS-2~\cite{Tisserand:2006zx} survey. The dot-dashed pink lines show the constraints from the dynamical effects on dwarf galaxies, from left to right: the p-value 0.01 constraint from mass segregation in Segue 1~\cite{Koushiappas:2017chw}, the tightest constraint from the disruption of the Eri II star cluster~\cite{Brandt:2016aco} and the weakest constraint from the disruption of ultra-faint dwarfs~\cite{Brandt:2016aco}. The short-dashed dark blue line shows the dynamical limit from the 25 most `halo like' wide binaries from Ref.~\cite{Monroy-Rodriguez:2014ula}. The short-long-dashed turquoise lines show the limits from the effects of radiation from primordial gas accreted onto PBHs in the early Universe on the Cosmic Microwave Background radiation, from left to right at low $M$: the constraint from Ref.~\cite{Horowitz:2016lib} and the tightest constraint from
Ref.~\cite{Ali-Haimoud:2016mbv} which assumes photoionization of gas (their limit assuming collisional ionization is significant weaker: $f=1$ is allowed for $M<10^{2} M_{\odot}$).
The long-dashed dark green lines show the limits from radio and X-ray emission due to accretion onto PBHs in the Milky Way, from left to right at low $M$: the $3\sigma$ X-ray constraint from Ref.~\cite{Gaggero:2016dpq} 
 the X-ray constraint, with no dark disc, from Ref.~\cite{Inoue:2017csr} and the $3\sigma$ radio constraint from Ref.~\cite{Gaggero:2016dpq}.}
\end{figure}

LIGO has detected gravitational waves from $\sim 10 \, M_{\odot} $ Black Hole (BH) binaries~\cite{Abbott:2016blz}. It has been suggested that these BHs could be PBHs which make up the DM~\cite{Bird:2016dcv,Clesse:2016vqa,Sasaki:2016jop}. The Milky Way halo fraction, $f$, of compact objects with mass $10^{-7} \lesssim  (M/M_{\odot}) \lesssim 10$ is tightly constrained by stellar microlensing~\cite{Paczynski:1985jf} observations of the Large Magellanic Cloud (LMC)~\cite{Allsman:2000kg,Tisserand:2006zx}. The dark matter fraction of more massive PBHs, $(M/M_{\odot}) \gtrsim 10$, is constrained by 
their dynamical effects on dwarf galaxies~\cite{Koushiappas:2017chw,Brandt:2016aco}
and halo wide binaries~\cite{Chaname:2003fn,Yoo:2003fr,Quinn:2009zg,Monroy-Rodriguez:2014ula}, X-ray and radio emission from accretion onto PBHs in the Milky Way~\cite{Gaggero:2016dpq,Inoue:2017csr} and the effects of radiation produced due to accretion onto PBHs in the early Universe on the Cosmic Microwave Background~\cite{Ricotti:2007au,Chen:2016pud,Blum:2016cjs,Ali-Haimoud:2016mbv,Horowitz:2016lib}.
See Fig.~\ref{fig:flimall} for a compilation of the constraints for $10^{-1} < (M/M_{\odot}) < 10^{2}$, assuming a delta-function mass function. Compact objects in this mass range will also microlens quasars~\cite{Chang:1979zz}. The observed variation in the brightness of images of multiply
imaged quasars is consistent with that expected from stars, hence limiting the contribution from other compact objects~\cite{Mediavilla:2017bok}. However constraints on the abundance of (dark) compact objects have not been calculated and therefore this constraint can not be included in  Fig.~\ref{fig:flimall}.

We see that, taken at face value, the constraints together exclude multi-Solar mass PBHs with a delta-function mass function making up all of the DM~\cite{Carr:2016drx}. Ref.~\cite{Carr:2016drx} argued that PBHs with an extended mass function (as produced by inflation models which have a broad peak in the primordial perturbation power spectrum) were consistent with all of the constraints. However Ref.~\cite{Green:2016xgy} showed that their method of applying constraints calculated assuming a delta-function mass function to extended mass functions was inaccurate, and that the quasi-log-normal mass functions produced by these inflation models were not consistent with all of the constraints in the multi-Solar mass range. Ref.~\cite{Kuhnel:2017pwq} subsequently studied the full range of possible PBH masses, and found a window around $(10^{-10}- 10^{-8}) M_{\odot}$ where PBHs with a quasi-log-normal mass function could make up all of the DM. However high-cadence microlensing observations of M31~\cite{Niikura:2017zjd} have subsequently placed tight constraints on this mass range. Recently Ref.~\cite{Carr:2017jsz} has considered a range of physically motivated extended mass functions. They found that the constraints are in fact tighter for extended mass function than for a delta-function mass function, and if all the constraints are taken at face value PBHs can not make up all of the DM.

All of these studies use the stellar microlensing limits calculated assuming a simple standard halo model. However it is known that the constraints depends significantly on the density and velocity distribution of the compact objects along the line of the sight to the LMC~\cite{Alcock:1995zx,Alcock:1996yv,Hawkins:2015uja}. In this paper we examine how uncertainties in the local density and circular speed affect the microlensing differential event rate and hence the constraints on both delta-function and quasi-log-normal mass functions. We also consider Evans' power law models~\cite{Evans} which allow for a non-flat rotation curve and hence a velocity distribution that varies with radius. In Sec.~\ref{sec:micro} we introduce these models and show how the microlensing differential event rate can vary. In Sec.~\ref{sec:constraints} we investigate how the constraints on both delta-function and quasi-log-normal mass functions change, before concluding with discussion in Sec.~\ref{sec:discussion}.

\section{Microlensing event rate}
\label{sec:micro}

Microlensing is the temporary amplification of a background star which occurs when a compact object passes close to the line of sight to the background star~\cite{Paczynski:1985jf}.
A microlensing event occurs when a compact object passes through the microlensing `tube', which has a radius of 
$u_{{\rm T}} R_{{\rm E}}$ where $u_{{\rm T}} \approx 1$ is the minimum impact
parameter for which the amplification of the background star is above
the required threshold and $R_{{\rm E}}$ is the Einstein radius:
\begin{equation}
R_{{\rm E}}(x)= 2 \left[ \frac{ G M x (1-x)L}{c^2 } \right]^{1/2} \,,
\end{equation}
where  $M$ is the mass of the compact object and $x$ is its distance from the observer in units of $L$, the distance to the source. For the LMC $L \approx 50 \, {\rm kpc}$.

Microlensing analyses usually assume a standard halo model (S), which consists of a cored isothermal sphere:
\begin{equation}
\rho(r) = \rho_{0} \frac{r_{{\rm c}}^2 + r_{0}^2}{r_{{\rm c}}^2 + r^2} \,,
\end{equation}
with local dark matter density $\rho_{0}= 0.008 M_{\odot} {\rm pc}^{-3}$, core radius $r_{{\rm c}} \approx 5$ kpc  and
Solar radius $r_{0} \approx 8.5$ kpc and an isotropic velocity distribution which is approximated to take the Maxwellian form
\begin{equation}
f(v) \, {\rm d}^3 v = \frac{1}{\left( \pi^{3/2} v_{\rm c}^3 \right)} \exp{\left( - \frac{v^2}{v_{\rm c}^2} \right)} \, {\rm d}^3 v \,,
\end{equation}
with local circular speed $v_{\rm c} = 220 \, {\rm km \, s}^{-1}$.

The differential event rate is then given by~\cite{Griest:1990vu,Alcock:1996yv}
\begin{eqnarray}
\frac{{\rm d} \Gamma}{{\rm d} \hat{t}} &=& \frac{512 \rho_{0} 
          (r_{{\rm c}}^2 + r_{0}^2) L G^2 u_{{\rm T}}^4 }
             {{\hat{t}}^4 {v_{{\rm c}}}^2 c^4}
            \nonumber \\ 
      \times   \int_{0}^{\infty} && \hspace{-0.5cm} \left[ \psi(M) M  
              \int^{x_{{\rm h}}\approx 1}_{0} \frac{x^2 (1-x)^2}{A + B x + x^2}
            e^{-Q(x) } {\rm d} x \right] {\rm d} M \,, 
\end{eqnarray}
where $\hat{t}$ is the time taken to cross the Einstein diameter, $Q(x)= 4 R^{2}_{{\rm E}}(x) u_{{\rm T}}^2 / (\hat{t}^{2} v_{{\rm c}}^2)$, $A=(r^2_{{\rm c}}+ r^2_{0})/L^2$, $B=-2(r_{0}/L) \cos{b}
\cos{l}$, $b=-33^{\circ}$ and $l=280^{\circ}$ are the Galactic
latitude and longitude respectively of the LMC and $\psi(M)$ is the mass function defined so that the fraction, $f$, of the total mass of the halo in the form of compact objects is
\begin{equation}
 f= \int_{0}^{\infty} \psi(M) \, {\rm d} M \,.
 \end{equation}

The expected number of events, $N_{\rm exp}$, is given by
\begin{equation}
N_{{\rm exp}} = E \int_{0}^{\infty} \frac{{\rm d} \Gamma}{{\rm d} \hat{t}}
           \,  \epsilon(\hat{t}) \, {\rm d} \hat{t} \,,
\end{equation}
where $E$ is the exposure in star years and $\epsilon(\hat{t})$ is the detection efficiency i.e. the probability that a microlensing event with duration $\hat{t}$ is detected. For the EROS-2 survey
$E=3.77 \times 10^{7}$ star years and the detection efficiency, in terms of Einstein radius crossing time, is given in Fig.~11 of Ref.~\cite{Tisserand:2006zx} (and as stated in the figure caption should be multiplied by a factor of 0.9 to take into account lensing by binary lenses). No events were observed and therefore 95\% confidence constraints on the fraction and mass function of compact objects can be calculated by requiring $N_{\rm exp} \leq 3.0$.

The resulting constraints on the halo fraction, $f$, of compact objects as a function of mass $M$,  from the EROS-2 survey~\cite{Tisserand:2006zx} and the MACHO collaboration $(1-30) M_{\odot}$ black hole search~\cite{Allsman:2000kg}, assuming a delta-function mass function, $\psi(M)=\delta(M)$, and model S,  are shown as dotted lines in Fig.~\ref{fig:flimall}. They are in good agreement with the published constraints (shown as solid lines). We subsequently only consider the EROS-2 survey, as it produces constraints which are the same as or tighter than the MACHO constraint in the mass region of interest.

We first consider the uncertainties in the parameters of halo model S. Determinations of the local dark matter density lie in the range $(0.005 - 0.015) \, M_{\odot} \, {\rm pc}^{-3}$ (or equivalently, in particle physics units, $(0.2-0.5) \, {\rm GeV} \, {\rm cm}^{-3}$)~\cite{Read:2014qva} and we consider the upper and lower limits of this range. We also consider the effects of a $10\%$ uncertainty in the local circular speed i.e. $v_{\rm c} = 220 \pm 20 \, {\rm km} \, {\rm s}^{-1}$~\cite{McMillan:2009yr}. In a specific halo model there is a one-to-one relationship between $v_{\rm c}$ and $\rho_{0}$, however we vary them individually, to assess their different effects on the differential event rate and the resulting constraints.

\begin{figure}[t]
%\begin{center}
\includegraphics[width=0.45\textwidth]{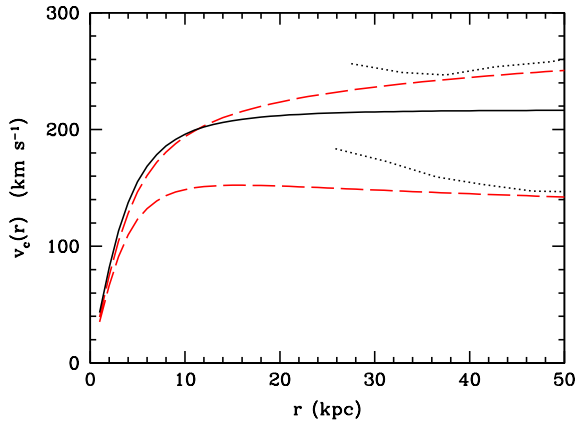}
%\end{center}
\caption{\label{fig:vc}The circular speed of the Milky Way, $v_{\rm c}(r)$, as a function of radius $r$. The solid black line is for the standard halo model S  and the long-dashed red lines for power law halos B and C (top and bottom respectively). The dotted black lines show the envelope of the compilation of observational data at large radii (where the halo dominates) from the right hand panel of Fig. 7 of Ref.~\cite{Bhattacharjee:2013exa}.}
\end{figure}

The microlensing differential event rate depends on the density and velocity distribution for $r_{0} < r< L$, therefore variations in these quantities with radius can have a significant effect on the event rate and the resulting constraints. This has been studied by the MACHO collaboration~\cite{Alcock:1995zx,Alcock:1996yv} using Evans' power law models~\cite{Evans} for which tractable expressions for the differential event rate exist~\cite{Alcock:1994qf}. These models have rotation curves at large radii ($r \gg r_{\rm c}$) which vary as $v_{\rm c}(r) \propto r^{-\beta}$ and also allow for flattening of the halo. We consider models `B' and `C' from Refs.~\cite{Alcock:1995zx,Alcock:1996yv}, which span the range of plausible models. Model B has a massive halo, total mass within 50 kpc $M(r<50 \, {\rm kpc})=7 \times 10^{11} \, M_{\odot}$, with a rising rotation curve ($\beta = -0.2$) with normalization velocity $v_{\rm a} = 200 \, {\rm km \, s}^{-1}$, while Model C has a light halo, $M(r<50 \, {\rm kpc})=2 \times 10^{11} \, M_{\odot}$, with a falling rotation curve ($\beta = 0.2$) and $v_{\rm a}= 180 \, {\rm km \, s}^{-1}$. Both models are spherical and have a core radius $r_{\rm c} = 5 \, {\rm kpc}$.  
For full details of the power law models see Refs.~\cite{Evans,Alcock:1994qf,Alcock:1995zx,Alcock:1996yv}. The lengthy expression for the differential event rate for the power law models is given in Appendix B of Ref.~\cite{Alcock:1994qf}.  

Fig.~\ref{fig:vc} shows the rotation curve, i.e. the variation of the circular speed, $v_{\rm c}(r)$, with radius
\begin{equation}
v_{\rm c}^2 (r)= v_{\rm a}^2 \frac{ r_{\rm c}^{\beta} r^2}{\left(r_{\rm c}^2 + r^2 \right)^{(\beta+2)/2} } \,,
\end{equation}
for these models, along with that of the standard halo S. We also plot the envelope of the compilation of observational data at large radii ($r>25 \, {\rm kpc}$), where the halo contribution to the rotation curve dominates, from the right hand panel of Fig. 7 of Ref.~\cite{Bhattacharjee:2013exa}. A detailed confrontation of the power law models with experimental data is beyond the scope of this work. However Fig.~\ref{fig:vc} demonstrates that these models are broadly consistent with recent data.

\begin{figure}[t]
%\begin{center}
\includegraphics[width=0.45\textwidth]{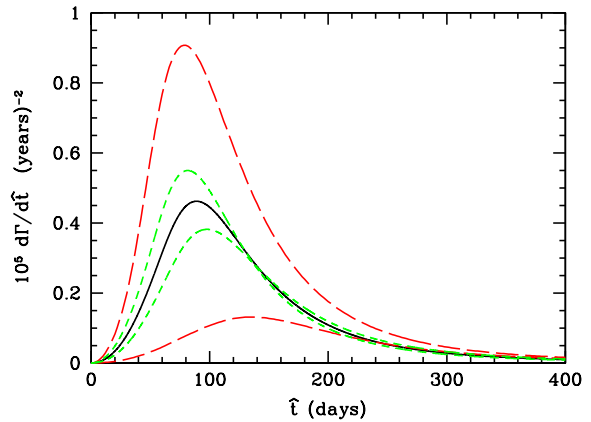}
%\end{center}
\caption{The LMC microlensing differential event rate as a function of Einstein diameter crossing time, $\hat{t}$, for compact objects with a delta-function mass function with $f=1$ and $M=1 M_{\odot}$. The solid black line is for model S with local circular speed $v_{\rm c} = 220 \, {\rm km \, s}^{-1}$, the short-dashed green lines are for  $v_{\rm c} = 240 \, {\rm km \, s}^{-1}$ and $200 \, {\rm km \, s}^{-1}$  (top and bottom respectively) and the long-dashed red lines for power law halos B and C (top and bottom respectively).}
\label{dgamdtfig}
\end{figure}

The LMC microlensing theoretical differential event rate (assuming detection efficiency $\epsilon(\hat{t})=1$) is shown in Fig.~\ref{dgamdtfig} for the standard halo S and cored isothermal spheres with local circular speeds of $v_{\rm c}= 200$ and $240 \, {\rm km \, s}^{-1}$ and for power law halo models B and C. Varying the local density, while keeping the circular speed fixed, only affects the normalisation of the differential event rate, so this is not shown in Fig.~\ref{dgamdtfig}. Changing the average velocity of the compact objects affects both the overall microlensing rate and the durations of the events. A smaller average velocity means that compact objects enter the microlensing tube less often, and hence the overall rate is smaller. They also spend more time within the microlensing tube and hence the typical event duration is increased. For a cored isothermal sphere the circular speed is independent of radius, and hence the velocity dispersion does not vary along the line of sight. In the power law halo models the circular speed, and hence the velocity dispersion, varies with radius and this leads to a greater variation in the microlensing differential event rates.

\begin{figure}[t]
%\begin{center}
\includegraphics[width=0.45\textwidth]{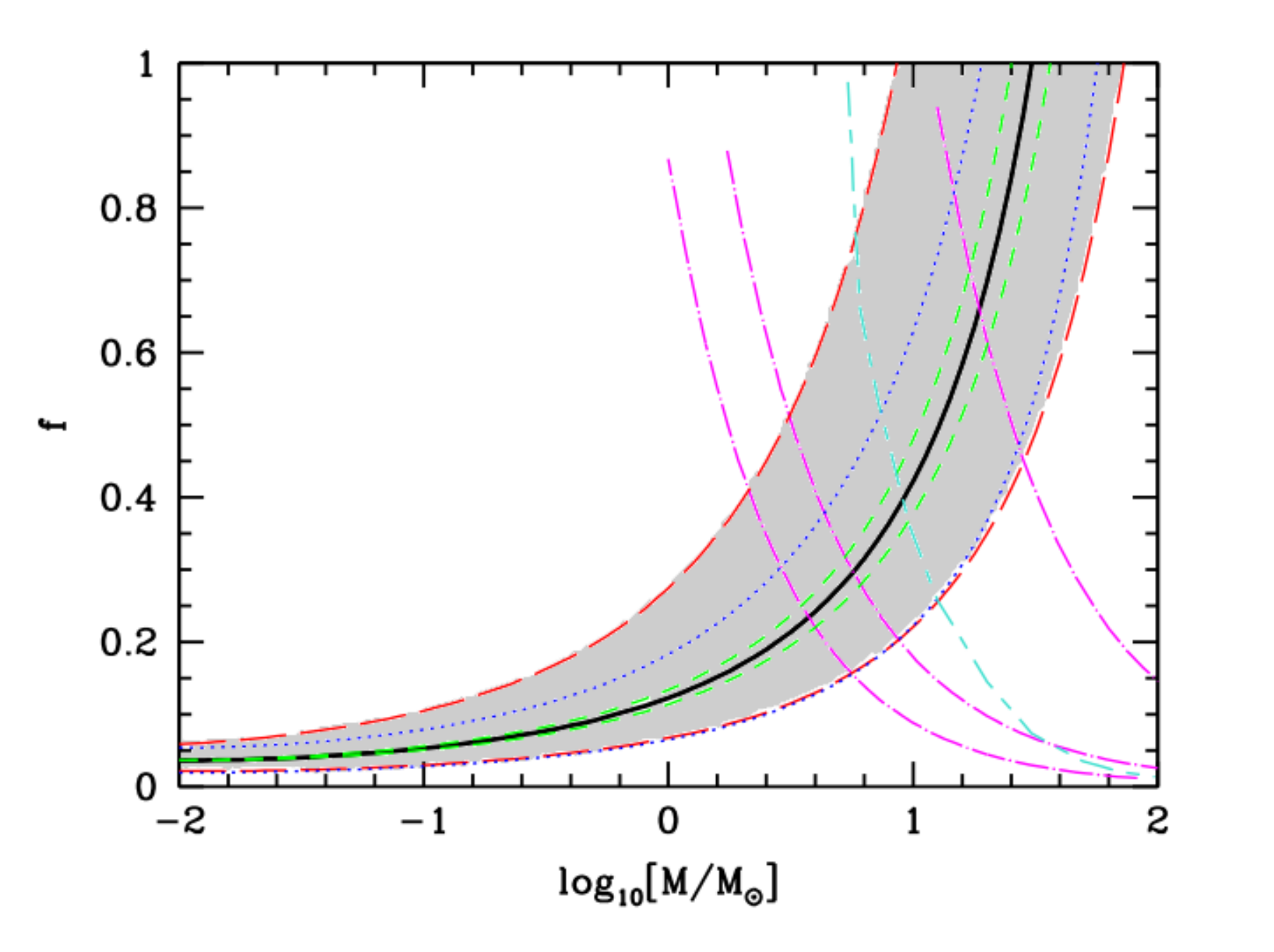}
%\end{center}
\caption{\label{flim} The dependence of the EROS-2 microlensing constraints on the halo fraction, $f$, on the modelling of the Milky Way halo for a delta-function mass function. The line types are as in Fig.~\ref{dgamdtfig}. The shaded region denotes the uncertainty in the microlensing constraint i.e. the difference between the tightest and the weakest constraint.
The tightest of the dynamical and accretion constraints, namely the dwarf mass segregation constraint~\cite{Koushiappas:2017chw} is also shown (left most dot-dashed pink line), along with the three constraints which have been calculated for a quasi-log-normal mass function: the tightest star cluster and weakest ultra-faint dwarf disruption limits~\cite{Brandt:2016aco} (dot-dashed pink lines) and the CMB constraint~\cite{Horowitz:2016lib} (short-long-dashed turquoise lines). }
\end{figure}

\section{Halo fraction constraints}
\label{sec:constraints}

Fig.~\ref{flim} shows the microlensing constraints from the EROS-2 survey on the halo fraction, $f$,  for the Milky Way halo models presented in Sec.~\ref{sec:micro}, assuming a delta-function mass function. The tightest of the dynamical and accretion limits, the dwarf mass segregation constraint from Ref.~\cite{Koushiappas:2017chw}, is also shown. For future reference we also show the the dynamical and accretion constraints that have been recalculated for a quasi-log-normal mass function, namely the tightest star cluster and weakest ultra-faint dwarf disruption limits from Ref.~\cite{Brandt:2016aco} and the CMB constraint from Ref.~\cite{Horowitz:2016lib}.

	\begin{table}
		  \begin{center}
		  \begin{ruledtabular}
		    \begin{tabular}{m{3.7cm} m{1.7cm}}
		    Halo model & $M_{\rm min}/M_{\odot}$ \\ \hline
  S  ($v_{\rm c} =220 \, {\rm km \, s}^{-1}$, $\rho_{0}= 0.008 M_{\odot} {\rm pc}^{-3}$) & 31 \\ \hline
   $v_{\rm c} =200 \, {\rm km \, s}^{-1}$ & 25 \\
  $v_{\rm c} =240 \, {\rm km \, s}^{-1}$ & 36 \\
  $\rho_{0} = 0.005 \, M_{\odot} \, {\rm pc}^{-3}$ & 19 \\
 $\rho_{0} = 0.015 \, M_{\odot} \, {\rm pc}^{-3}$ & 57 \\
 B  & 73 \\
 C & 8.7 \\ 
		\end{tabular}
		\end{ruledtabular}
		  \end{center}
		  \caption{The smallest value of $M$ for which a delta-function mass function with $f=1$ is consistent with the EROS microlensing observations, $M_{\rm min}$, for the halo models presented in Sec.~\ref{sec:micro}. See text for details of models.}
		\label{tab}
		\end{table}

Table~\ref{tab} gives the smallest value of $M$ for which a delta-function mass function with $f=1$ is consistent with the EROS-2 microlensing observations, $M_{\rm min}$, for each halo model. The effect of varying the local density alone is straightforward; increasing the density, increases the microlensing event rate and hence increases $M_{\rm min}$.
The models with lower velocity dispersion (i.e. the cored isothermal sphere with $v_{\rm c} = 200 \, {\rm km \, s}^{-1}$ and power law model C) have smaller event rates, so the constraints are weakened and $M_{\rm min}$ is smaller. For the lower (upper) limit on the local density
$\rho_{\rm 0} = 0.005 \, (0.015) \, M_{\odot} \, {\rm pc}^{-3}$ $M_{\rm min}$ is decreased by $\sim 40\%$ (increased by $\sim 80\%$). Varying the local circular speed, $v_{\rm c}$, by $10\%$, while keeping the local density fixed,  changes $M_{\rm min}$ by $\sim 20\%$. $M_{\rm min}$ is increased by a factor of 2 (decreased by 3.5) for model B (C). For model C $M_{\rm min}= 9 M_{\odot} $. This is slightly smaller than the largest mass for which a delta-function mass function with $f=1$ is allowed by the weakest dwarf disruption constraint, $M_{\rm max}= 12 M_{\odot}$~\cite{Brandt:2016aco} i.e.~a delta-function mass function with $f=1$ and $M \approx 10 M_{\odot}$ is compatible with both these constraints. However there are other tighter dynamical constraints in this mass range;  the disruption of the star cluster in Eri II~\cite{Brandt:2016aco} and mass segregation in Segue 1~\cite{Koushiappas:2017chw} both have $M_{\rm max} \approx 1 M_{\odot}$.

We now apply the microlensing constraint to extended mass functions. A log-normal mass function is a good fit to the mass functions produced by inflation models with a broad peak in the primordial power spectrum~\cite{Green:2016xgy,Carr:2017jsz}.
For computational convenience, as in Ref.~\cite{Green:2016xgy}, we use a quasi-log-normal mass function where the pre-factor multiplying the exponential is independent of mass:
\begin{equation}
\label{lnmf}
\psi(M)  =  N \exp{ \left\{ -  \frac{ \left[ \log{(M/M_{\odot})} - \log{(M_{\rm c}}/M_{\odot})\right]^2}{2 \sigma^2} \right\}} \,,
\end{equation}
where $N$ is a normalisation constant that we fix so that the halo fraction is normalised to unity.

\begin{figure}[t]
%\begin{center}
\includegraphics[width=0.45\textwidth]{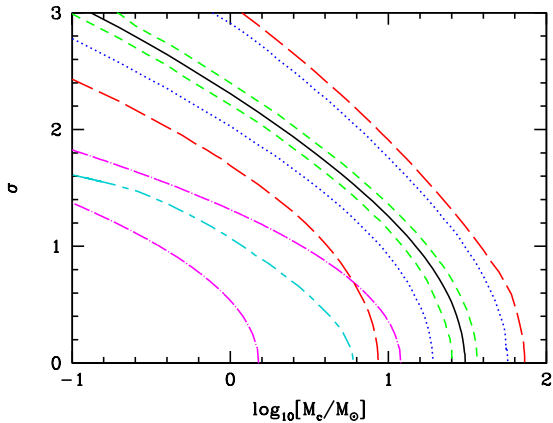}
%\end{center}
\caption{\label{flim2} Constraints on the width, $\sigma$, of the quasi-log-normal mass function, Eq.~(\ref{lnmf}), as a function of the central mass, $M_{\rm c}$, from the EROS-2 microlensing survey. The line types are as in Figs.~\ref{dgamdtfig} and \ref{flim} and for the microlensing constraints parameters beneath the lines are excluded. The tightest star cluster and weakest ultra-faint dwarf disruption limits from Ref.~\cite{Green:2016xgy} and the CMB constraint from Ref.~\cite{Horowitz:2016lib} are also shown. For these constraints the areas above the lines are excluded.}
\end{figure}

Fig.~\ref{flim2} shows the $\sigma$ and $M_{\rm c}$ values excluded by the EROS-2 microlensing survey for the halo models presented in Sec.~\ref{sec:micro} and studied above for a delta-function mass function. We also show the dynamical and accretion constraints that have been calculated for a quasi-log-normal mass function, namely the tightest star cluster and weakest ultra-faint dwarf disruption limits~\cite{Green:2016xgy}, calculated using the same prescriptions as used in Ref.~\cite{Brandt:2016aco} for delta-function mass functions, and the CMB constraints~\cite{Horowitz:2016lib}. Tighter limits are expected, from microlensing of quasars~\cite{Mediavilla:2017bok} and mass segregation in dwarf galaxies~\cite{Koushiappas:2017chw}. However constraints on $f$ have not been calculated for the former and for the later they have not been recalculated for extended mass functions. The regions of parameter space excluded by the EROS-2 microlensing survey vary significantly; for a given central mass $M_{\rm c}$, the maximum allowed value of $\sigma$ for model C is roughly twice as large as for model B. However the general tension between the microlensing constraints and the dynamical and accretion constraints remains. 
There is a small region of parameter space, $M_{\rm c} \sim 10 M_{\odot}$ and small $\sigma$, which is consistent with both the stellar microlensing constraint for model C and also the weakest dwarf galaxy disruption limit. The existence of this region is expected, since a delta-function mass function with $M_{\rm c} \sim 10 M_{\odot}$ was consistent with both these constraints. However both constraints are only satisfied for $\sigma < 0.6$ i.e.~increasing the width of the mass function  increases 
the tension between the stellar microlensing and dynamical constraints. This confirms the result recently found in Ref.~\cite{Carr:2017jsz} for a wider range of mass functions and constraints.

\section{Discussion}
\label{sec:discussion}

Stellar microlensing constrains the halo fraction of compact objects with $10^{-7} <M /M_{\odot} < 10$, while dynamical and accretion constraints constrain the abundance of PBHs with $M/M_{\odot} \gtrsim 10$. Taken at face value together they exclude PBHs with $10^{-7} <M /M_{\odot} < 10^{5}$ making up all of the dark matter~\cite{Carr:2016drx,Green:2016xgy,Kuhnel:2017pwq,Carr:2017jsz}. However the microlensing differential event rate, and hence the resulting constraints on compact objects, depend on their density and velocity distribution along the line of sight to the LMC.

We have studied how the constraints are affected by astrophysical uncertainties. We first varied the parameters of the standard halo model, S, used in microlensing studies, a cored isothermal sphere with an isotropic Maxwellian velocity distribution. The differential event rate is directly proportional to the local density, whereas varying the local circular speed affects both the total rate and the durations of the events. We then turned to Evans' power law models where the circular speed can vary with radius~\cite{Evans}. We looked at two specific models which have been used by the MACHO microlensing collaboration~\cite{Alcock:1995zx,Alcock:1996yv} and are broadly consistent with observations of the rotation curve of the Milky Way~\cite{Bhattacharjee:2013exa}. Model B has a massive halo with a rising rotation curve, while model C has a light halo with a falling rotation curve.

For the standard halo model $M_{\rm min}$, the smallest mass for which a delta-function mass function with $f=1$ is allowed by the EROS-2 survey, is $31 M_{\odot}$. Varying the local circular speed, $v_{\rm c}$, by $10\%$, while keeping the local density fixed,  changes  $M_{\rm min}$ by $\sim 20\%$. For local densities in the range $\rho_{\rm 0} = (0.005-0.015) \, M_{\odot} \, {\rm pc}^{-3}$ $M_{\rm min}$ lies between $19$ and $57 M_{\odot}$ for fixed $v_{\rm c}$. Models B and C, where the velocity distribution varies with radius, have larger changes in $M_{\rm min}$: for model B (C) $M_{\rm min} = 83 \, (8.7) \, M_{\odot}$. The value of  $M_{\rm min}$ for model C is slightly smaller than the largest mass for which a delta-function mass function with $f=1$ is allowed by the weakest dwarf disruption constraint, $M_{\rm max}=12 M_{\odot}$~\cite{Brandt:2016aco}. In other words a delta-function mass function with $f=1$ and $M \approx 10 M_{\odot}$ is compatible with both the microlensing constraint for a light halo and the weakest dwarf galaxy disruption constraint. However there are tighter constraints on compact objects in this mass range from the disruption of the star cluster in Eri II~\cite{Brandt:2016aco}, mass segregation in Segue 1~\cite{Koushiappas:2017chw} and potentially also the microlensing of quasars~\cite{Mediavilla:2017bok}.

We then looked at the constraints on quasi-log-normal mass functions, which are produced by inflation models with a broad feature in the primordial power spectrum. For a given central mass $M_{\rm c}$, the maximum allowed value of the width $\sigma$ is roughly twice as large for model C as it is for model B. There is a small region of parameter space, with $M_{\rm c} \sim 10 M_{\odot}$ and small $\sigma$, which is consistent with both the stellar microlensing constraint for the light halo model C and also the weakest dwarf galaxy disruption limit. However, as recently found in Ref.~\cite{Carr:2017jsz}, the tension between the constraints is in fact increased relative to the case of the delta-function mass function. 

In summary, astrophysical uncertainties have a non-negligible effect on the constraints on PBHs, and other compact objects, from stellar microlensing observations. This effect is unlikely to be large enough to reconcile the microlensing constraints with the current dynamical constraints and allow all of the DM to be in the form of multi-Solar mass PBHs. However this illustrates the importance of taking into account astrophysical uncertainties and assumptions when calculating and comparing constraints on PBH DM.

\vspace*{1cm}
\acknowledgments

A.M.G.  acknowledges  support  from  STFC  grant ST/L000393/1.


\begin{thebibliography}{1}


\bibitem{Carr:1974nx}
  B.~J.~Carr and S.~W.~Hawking,
  %``Black holes in the early Universe,''
  Mon.\ Not.\ Roy.\ Astron.\ Soc.\  {\bf 168} (1974) 399.

\bibitem{Carr:1975qj}
  B.~J.~Carr,
  %``The Primordial black hole mass spectrum,''
  Astrophys.\ J.\  {\bf 201} (1975) 1.
%  doi:10.1086/153853


\bibitem{Carr:2016drx}
  B.~Carr, F.~Kuhnel and M.~Sandstad,
  %``Primordial Black Holes as Dark Matter
 arXiv:1607.06077 [astro-ph.CO].

\bibitem{Abbott:2016blz}
  B.~P.~Abbott {\it et al.} [LIGO Scientific and Virgo Collaborations],
  %``Observation of Gravitational Waves from a Binary Black Hole Merger,''
  Phys.\ Rev.\ Lett.\  {\bf 116} (2016) no.6,  061102
  %doi:10.1103/PhysRevLett.116.061102
  [arXiv:1602.03837 [gr-qc]].


\bibitem{Bird:2016dcv}
  S.~Bird, I.~Cholis, J.~B.~Muñoz, Y.~Ali-Haïmoud, M.~Kamionkowski, E.~D.~Kovetz, A.~Raccanelli and A.~G.~Riess,
  %``Did LIGO detect dark matter?,''
  Phys.\ Rev.\ Lett.\  {\bf 116} (2016) no.20,  201301
  %doi:10.1103/PhysRevLett.116.201301
  [arXiv:1603.00464 [astro-ph.CO]].

\bibitem{Clesse:2016vqa}
  S.~Clesse and J.~García-Bellido,
  %``The clustering of massive Primordial Black Holes as Dark Matter: measuring their mass distribution with Advanced LIGO,''
  arXiv:1603.05234 [astro-ph.CO].


\bibitem{Sasaki:2016jop}
  M.~Sasaki, T.~Suyama, T.~Tanaka and S.~Yokoyama,
  %``Primordial Black Hole Scenario for the Gravitational-Wave Event GW150914,''
  Phys.\ Rev.\ Lett.\  {\bf 117} (2016) no.6,  061101
 % doi:10.1103/PhysRevLett.117.061101
  [arXiv:1603.08338 [astro-ph.CO]].





  
  \bibitem{Paczynski:1985jf}
  B.~Paczynski,
  %``Gravitational microlensing by the galactic halo,''
  Astrophys.\ J.\  {\bf 304} (1986) 1.
 % doi:10.1086/164140



\bibitem{Allsman:2000kg}
  R.~A.~Allsman {\it et al.} [Macho Collaboration],
  %``MACHO project limits on black hole dark matter in the 1-30 solar mass range,''
  Astrophys.\ J.\  {\bf 550} (2001) L169
  %doi:10.1086/319636
  [astro-ph/0011506].

\bibitem{Tisserand:2006zx}
  P.~Tisserand {\it et al.} [EROS-2 Collaboration],
  %``Limits on the Macho Content of the Galactic Halo from the EROS-2 Survey of the Magellanic Clouds,''
  Astron.\ Astrophys.\  {\bf 469} (2007) 387
 % doi:10.1051/0004-6361:20066017
  [astro-ph/0607207].

\bibitem{Koushiappas:2017chw}
  S.~M.~Koushiappas and A.~Loeb,
  %``Dynamics of dwarf galaxies disfavor stellar-mass black hole dark matter,''
  arXiv:1704.01668 [astro-ph.GA].

\bibitem{Brandt:2016aco}
  T.~D.~Brandt,
  %``Constraints on MACHO Dark Matter from Compact Stellar Systems in Ultra-Faint Dwarf Galaxies,''
  Astrophys.\ J.\  {\bf 824} (2016) no.2,  L31
 % doi:10.3847/2041-8205/824/2/L31
  [arXiv:1605.03665 [astro-ph.GA]].


\bibitem{Chaname:2003fn}
  J.~Chaname and A.~Gould,
  %``Disk and halo wide binaries from the revised luyten catalog: probes of star formation and MACHO dark matter,''
  Astrophys.\ J.\  {\bf 601} (2004) 289
  %doi:10.1086/380442
  [astro-ph/0307434].

\bibitem{Yoo:2003fr}
  J.~Yoo, J.~Chaname and A.~Gould,
  %``The end of the MACHO era: limits on halo dark matter from stellar halo wide binaries,''
  Astrophys.\ J.\  {\bf 601} (2004) 311
  %doi:10.1086/380562
  [astro-ph/0307437].


\bibitem{Quinn:2009zg}
  D.~P.~Quinn, M.~I.~Wilkinson, M.~J.~Irwin, J.~Marshall, A.~Koch and V.~Belokurov,
  %``On the Reported Death of the MACHO Era,''
  Mon.\ Not.\ Roy.\ Astron.\ Soc.\  {\bf 396} (2009) 11
  %doi:10.1111/j.1745-3933.2009.00652.x
  [arXiv:0903.1644 [astro-ph.GA]].

\bibitem{Monroy-Rodriguez:2014ula}
  M.~A.~Monroy-Rodríguez and C.~Allen,
  %``The end of the MACHO era- revisited: new limits on MACHO masses from halo wide binaries,''
  Astrophys.\ J.\  {\bf 790} (2014) no.2,  159
  %doi:10.1088/0004-637X/790/2/159
  [arXiv:1406.5169 [astro-ph.GA]].

\bibitem{Gaggero:2016dpq}
  D.~Gaggero, G.~Bertone, F.~Calore, R.~M.~T.~Connors, M.~Lovell, S.~Markoff and E.~Storm,
  %``Searching for Primordial Black Holes in the radio and X-ray sky,''
  arXiv:1612.00457 [astro-ph.HE].
  
  \bibitem{Inoue:2017csr}
  Y.~Inoue and A.~Kusenko,
  %``A new X-ray bound on primordial black holes density,''
  arXiv:1705.00791 [astro-ph.CO].


\bibitem{Ricotti:2007au}
  M.~Ricotti, J.~P.~Ostriker and K.~J.~Mack,
  %``Effect of Primordial Black Holes on the Cosmic Microwave Background and Cosmological Parameter Estimates,''
  Astrophys.\ J.\  {\bf 680} (2008) 829
 % doi:10.1086/587831
  [arXiv:0709.0524 [astro-ph]].

\bibitem{Chen:2016pud} 
  L.~Chen, Q.~G.~Huang and K.~Wang,
  %``Constraint on the abundance of primordial black holes in dark matter from Planck data,''
  arXiv:1608.02174 [astro-ph.CO].


\bibitem{Blum:2016cjs}
  D.~Aloni, K.~Blum and R.~Flauger,
  %``CMB constraints on primordial black hole dark matter,''
  JCAP {\bf 1705} (2017) no.05,  017
  %doi:10.1088/1475-7516/2017/05/017
  [arXiv:1612.06811 [astro-ph.CO]].


\bibitem{Ali-Haimoud:2016mbv}
  Y.~Ali-Haïmoud and M.~Kamionkowski,
  %``Cosmic microwave background limits on accreting primordial black holes,''
  Phys.\ Rev.\ D {\bf 95} (2017) no.4,  043534
  %doi:10.1103/PhysRevD.95.043534
  [arXiv:1612.05644 [astro-ph.CO]].


\bibitem{Horowitz:2016lib}
  B.~Horowitz,
  %``Revisiting Primordial Black Holes Constraints from Ionization History,''
  arXiv:1612.07264 [astro-ph.CO].
  

  
  
  
  
\bibitem{Chang:1979zz}
  K.~Chang and S.~Refsdal,
  %``Flux variations of QSO 0957+561 A, B and image splitting by stars near the light path,''
  Nature {\bf 282} (1979) 561.
  %doi:10.1038/282561a0
  
  
  \bibitem{Mediavilla:2017bok}
  E.~Mediavilla, J.~Jiménez-Vicente, J.~A.~Muñoz, H.~Vives-Arias and J.~Calderón-Infante,
  %``Limits on the Mass and Abundance of Primordial Black Holes from Quasar Gravitational Microlensing,''
  Astrophys.\ J.\  {\bf 836} (2017) no.2,  L18
 % doi:10.3847/2041-8213/aa5dab
  [arXiv:1702.00947 [astro-ph.GA]].





\bibitem{Green:2016xgy}
  A.~M.~Green,
  %``Microlensing and dynamical constraints on primordial black hole dark matter with an extended mass function,''
  Phys.\ Rev.\ D {\bf 94} (2016) no.6,  063530
  %doi:10.1103/PhysRevD.94.063530
  [arXiv:1609.01143 [astro-ph.CO]].


\bibitem{Kuhnel:2017pwq}
  F.~Kuhnel and K.~Freese,
  %``Constraints on Primordial Black Holes with Extended Mass Functions,''
  Phys.\ Rev.\ D {\bf 95} (2017) no.8,  083508
  %doi:10.1103/PhysRevD.95.083508
  [arXiv:1701.07223 [astro-ph.CO]].



\bibitem{Niikura:2017zjd}
  H.~Niikura {\it et al.},
  %``Microlensing constraints on $10^{-10}M_\odot$-scale primordial black holes from high-cadence observation of M31 with Hyper Suprime-Cam,''
  arXiv:1701.02151 [astro-ph.CO].

\bibitem{Carr:2017jsz}
  B.~Carr, M.~Raidal, T.~Tenkanen, V.~Vaskonen and H.~Veermäe,
  %``Primordial black hole constraints for extended mass functions,''
  arXiv:1705.05567 [astro-ph.CO].


 \bibitem{Alcock:1995zx}
  C.~Alcock {\it et al.} [MACHO Collaboration],
  %``The MACHO project first year LMC results: The Microlensing rate and the nature of the galactic dark halo,''
  Astrophys.\ J.\  {\bf 461} (1996) 84
  %doi:10.1086/177039
  [astro-ph/9506113].
  
  \bibitem{Alcock:1996yv}
  C.~Alcock {\it et al.} [MACHO Collaboration],
  %``The MACHO project LMC microlensing results from the first two years and the nature of the galactic dark halo,''
  Astrophys.\ J.\  {\bf 486} (1997) 697
  %doi:10.1086/304535
  [astro-ph/9606165].  


\bibitem{Hawkins:2015uja}
  M.~R.~S.~Hawkins,
  %``A new look at microlensing limits on dark matter in the Galactic halo,''
  Astron.\ Astrophys.\  {\bf 575} (2015) A107
  %doi:10.1051/0004-6361/201425400
  [arXiv:1503.01935 [astro-ph.GA]].


\bibitem{Evans}
 N.~W.~Evans,
  Mon.\ Not.\ Roy.\ Astron.\ Soc.\  {\bf 267} (1994) 333.








\bibitem{Griest:1990vu}
  K.~Griest,
  %``Galactic Microlensing as a Method of Detecting Massive Compact Halo Objects,''
  Astrophys.\ J.\  {\bf 366} (1991) 412.
 % doi:10.1086/169575



\bibitem{Read:2014qva}
  J.~I.~Read,
  %``The Local Dark Matter Density,''
  J.\ Phys.\ G {\bf 41} (2014) 063101
 % doi:10.1088/0954-3899/41/6/063101
  [arXiv:1404.1938 [astro-ph.GA]].

\bibitem{McMillan:2009yr}
  P.~J.~McMillan and J.~J.~Binney,
  %``The uncertainty in Galactic parameters,''
  Mon.\ Not.\ Roy.\ Astron.\ Soc.\  {\bf 402} (2010) 934
  %doi:10.1111/j.1365-2966.2009.15932.x
  [arXiv:0907.4685 [astro-ph.GA]].







\bibitem{Alcock:1994qf}
  C.~Alcock {\it et al.} [MACHO Collaboration],
  %``Theory of exploring the dark halo with microlensing. 1: Power law models,''
  Astrophys.\ J.\  {\bf 449} (1995) 28
  %[astro-ph/9411019].


   
  \bibitem{Bhattacharjee:2013exa}
  P.~Bhattacharjee, S.~Chaudhury and S.~Kundu,
  %``Rotation Curve of the Milky Way out to $\sim$ 200 kpc,''
  Astrophys.\ J.\  {\bf 785} (2014) 63
  [arXiv:1310.2659 [astro-ph.GA]].
  
\end{thebibliography}
\end{document}